\documentclass[twoside,12pt]{article}
\usepackage{epsf,epsfig,amssymb,amsmath,graphicx,float}
\usepackage{color}
\usepackage{hyperref}
\usepackage{indentfirst}
\hypersetup{colorlinks,citecolor=blue,linkcolor=red,urlcolor=red}

\usepackage{amsmath}

\newcommand{\gsim}{\raisebox{-0.13cm}{~\shortstack{$>$ \\[-0.07cm] $\sim$}}~}

\begin{document}
\renewcommand{\thefootnote}{\fnsymbol{footnote}}

\begin{titlepage}

\begin{center}

\vspace{1cm}

{\Large {\bf Constraints on Asymmetric Dark Matter Self Annihilation Cross Sections  }}

\vspace{1cm}

{\bf Qiquan Li\textsuperscript{1}, Hoernisa Iminniyaz\textsuperscript{1,\footnote{Corresponding author, wrns@xju.edu.cn}}, Fangyu Liu\textsuperscript{2}}

\vskip 0.15in
{\it
\textsuperscript{1}School of Physics Science and Technology, Xinjiang University, \\
Urumqi 830017, China \\[0.5em]
\textsuperscript{2}Tsung-Dao Lee Institute, Shanghai Jiao Tong University, \\
Shanghai 201210, China
}

\abstract{ We investigate the evolution of the relic density for asymmetric dark matter incorporating self annihilation processes in both the shear-dominated universe and Gauss-Bonnet braneworld. Under the same conditions where the ratio of final asymmetry to the initial asymmetry $ R \equiv \Delta_{-\infty}/\Delta_{\text{in}} $ is identical, the shear-dominated universe, due to its enhanced Hubble expansion rate, leads to an earlier freeze-out point of wash-out asymmetry process and allows a higher upper limit on the self annihilation cross section. Conversely, the Gauss-Bonnet braneworld, with a weakened Hubble expansion rate, delays the freeze-out point and permits a lower upper limit on the self annihilation cross section. We further constrain the wino mass $M_2$ for sneutrino and higgsino asymmetric dark matter in both scenarios, showing that, compared to the standard model, the lower limit of $M_2$ is smaller in the shear-dominated universe but higher in the Gauss-Bonnet braneworld.
 }
\end{center}
\end{titlepage}
\setcounter{footnote}{0}

\section{Introduction}
According to Planck 2018 observations of the cosmic microwave background (CMB) anisotropies \cite{Planck:2018vyg}, the relic density of the present cold dark matter is measured as
\begin{equation}
    \Omega_{\mathrm{CDM}} h^2 = 0.120 \pm 0.001 \quad (68\%\ \text{C.L.}),
\end{equation}
where $h = 0.674 \pm 0.005$ is the present Hubble expansion rate in units of $100\ \mathrm{km\,s^{-1}\,Mpc^{-1}}$ \cite{Planck:2018vyg}. This result, combined with multi-probe astrophysical observations including galaxy rotation curves \cite{Pato:2015dua,Iocco:2015xga,Jiao:2023aci,Ou:2023adg}, gravitational lensing \cite{Clowe:2006eq,Robertson:2016qef}, and large-scale structure \cite{Springel:2005nw}, establishes robust evidence for dark matter's existence.

The identity of dark matter particles are considered, as there are no particles in the Standard Model (SM) that exhibit the properties of the best motivated candidates, such as neutral, long-lived, or stable Weakly Interacting Massive Particles (WIMPs). To address this, physicists have proposed various candidates beyond the SM frameworks. For instance, supersymmetric models predict neutral Majorana particles where particles and anti-particles are identical \cite{Bertone:2004pz}. Nevertheless, the symmetry of dark matter remains uncertain, this making Asymmetric Dark Matter (ADM) a viable candidate as well.

The ADM model proposes that there exists an asymmetry between dark matter particles $\chi$ and anti-particles $\bar\chi$, similar to the baryon asymmetry in the universe, where the number density of dark matter particles significantly exceeds anti-particles \cite{Nussinov:1985xr,Chivukula:1989qb,An:2009vq,Cohen:2009fz,Shelton:2010ta}. The asymmetry of dark matter may originate from coupling mechanisms with baryon/lepton asymmetries in the early Universe (e.g., sphaleron processes), where the CP violation induced asymmetry is transferred under thermal equilibrium conditions and subsequently preserved after freeze-out \cite{Hooper:2004dc,Abel:2006nv,Kaplan:2009ag,XENON100:2011uwh}.

The evolution of the relic density of ADM is influenced by factors such as the annihilation rate of dark matter particles and anti-particles, the self annihilation rate, and the Hubble expansion rate \cite{Kolb}. In most studies, only the annihilation rate and the Hubble expansion rate are considered. In the early high-temperature, high-density universe, dark matter particles and anti-particles maintain thermal equilibrium through annihilation and inverse reactions. During this phase, the particle number density is temperature dominated, with the annihilation rate significantly exceeding the Hubble expansion rate. As the universe expands and cools, when the temperature approaches the ADM scale $T \sim m$, the annihilation rate falls below Hubble expansion rate. The particles and anti-particles then decouple from thermal equilibrium, and they enter a freeze-out of ADM number density state, with the co-moving ADM number density remaining nearly constant.

Notably, the discrete $\mathbb{Z}_{N}$ symmetries (e.g., $R$-parity) allow self annihilation processes. If we consider the self annihilation of particles and anti-particles in the above mentioned evolution of ADM, the asymmetry between dark matter particles and anti-particles may be washed out. The earliest consideration of the self annihilation of ADM particles and anti-particles in the standard cosmological scenario was by Ellwanger and Mitropoulos \cite{Ellwanger:2012yg}. They found that when the temperature approaches the dark matter mass $T \sim m$ and the rate of ADM self annihilation is lower than the Hubble expansion rate, the asymmetry is not significantly washed out \cite{Ellwanger:2012yg}. Subsequent studies by Liu and Iminniyaz extended this analysis to kination model and brane world cosmology \cite{Liu:2023pfd}.

Although the standard cosmological model $\Lambda$CDM has achieved remarkable success, its description of the early universe evolution still lacks conclusive evidence. In response to this limitation, the relic density of ADM in several non-standard cosmological scenarios was investigated in \cite{Iminniyaz:2013cla,Meehan:2014zsa,Abdusattar:2015azp,Meehan:2015cna}. The shear-dominated universe, based on the Bianchi type I anisotropic spacetime solution, assumes that the early universe expansion was dominated by shear energy density $\rho_s \propto \bar{R}^{-6}$ where $ \bar{R}$ is the mean-scale factor of the universe, leading to a non-standard Hubble expansion rate evolution $H \propto \bar{R}^{-3}$ prior to the radiation dominated era. The relic abundance of dark matter in the Bianchi type I cosmological model was investigated in \cite{Barrow:1982ei,Kamionkowski:1990ni}, and further studied for ADM within this model in \cite{Iminniyaz:2016iom}. On the other hand, the Gauss-Bonnet braneworld model originates from higher-dimensional gravitational theory \cite{Langlois:2002bb}. It assumes that standard model particles are confined on a 3-brane, while introducing the Gauss-Bonnet curvature term from string theory to modify the gravitational field equations \cite{Okada:2009xe}. This leads to a Hubble expansion rate evolution $H \propto \rho^{1/3}$ in high-energy regimes. The relic abundances of both DM and ADM in the Gauss-Bonnet braneworld were studied in \cite{Okada:2009xe,Meehan:2014bya}. The above study on ADM in a shear-dominated universe and Gauss-Bonnet braneworld only considered the annihilation process between particles and anti-particles, excluding self annihilation. In this paper, we investigate the evolution of the relic density of ADM using Boltzmann equations in both scenarios that include self annihilation processes. We solve the coupled set of Boltzmann equations including the self-annihilation process in the shear-dominated universe and the Gauss-Bonnet braneworld. It is found that the enhanced and weakened cosmic expansion rates lead to an earlier and delay the freeze-out point of the wash-out asymmetry process, respectively. Consequently, the upper limits of the self annihilation cross section for the final asymmetry in the shear-dominated universe and Gauss-Bonnet braneworld are higher and lower than those in the standard cosmological scenario, respectively.

The structure of this paper is as follows. In Sec.~\ref{sec:2}, we first present the Boltzmann equations in non-standard cosmological models. The evolution of the relic abundance of ADM and constraints on the self annihilation cross sections of ADM particles and antiparticles are analyzed in Section~\ref{sec:3}. In Sec.~\ref{sec:4}, we constrain the wino mass for sneutrino and higgsino ADM in a shear-dominated universe and Gauss-Bonnet braneworld. Finally, the conclusion and summary are presented in Sec.~\ref{sec:5}.
\section{Relic abundance of ADM in shear-dominated universe and Gauss-Bonnet braneworld }
\subsection{Boltzmann equations in non-standard cosmological scenarios}\label{sec:2}
In this subsection, we consider the modification of the Hubble expansion rate $H$ under the shear-dominated universe and Gauss-Bonnet braneworld, and further present the Boltzmann equations for ADM particles and anti-particles involving self annihilation processes in the non-standard cosmological scenarios. The Hubble expansion rate in a shear-dominated universe is \cite{Barrow:1982ei,Kamionkowski:1990ni}
\begin{align} \label{eq:B1}
H_{s} = H\sqrt{ 1 + \frac{x_e^2}{x^2} },
\end{align}
where $H = \pi \sqrt{g_*/90} T^2/M_{\text{Pl}}$ is the Hubble expansion rate in standard cosmological scenario with $M_{\text{Pl}} = 2.4 \times 10^{18}\ \mathrm{GeV}$ being the reduced Planck mass, and $g_*$ is the effective number of relativistic degrees of freedom. Here $x = m/T$ is the mass to temperature ratio, $x_e = \sqrt{g_*/g_*^\text{e}} m/T_e$ is the shear factor, and $g_*^\text{e}$ denotes the value of $g_*$ at $T_e$. The Hubble expansion rate in the Gauss-Bonnet braneworld is given by \cite{Meehan:2014bya}
\begin{align} \label{eq:B2}
H_{g}=H \left( \frac{x}{x_t} \right)^{2/3},
\end{align}
where $x_t = m_\chi/T_t$ and $T_t$ is the transition temperature. The modified expansion rate transitions to the standard cosmological expansion rate when the temperature drops below $T_t$.

The Boltzmann equation is used to calculate the relic abundance of ADM particles and anti-particles. We consider the Boltzmann equation for ADM particles and anti-particles involving self annihilation processes in the shear-dominated universe and Gauss-Bonnet braneworld as follows:
\begin{eqnarray} \label{eq:B3}
  \frac{{\rm d}n_{\chi,\bar\chi}}{{\rm d}t} + 3 H_{s,g} n_{\chi,\bar\chi} =
       - \langle \sigma_{\chi\bar\chi} v\rangle
        (n_{\chi} n_{\bar\chi} - n^{\rm eq}_{\chi} n^{\rm eq}_{\bar\chi})\,
        - \langle \sigma_{\chi\chi} v\rangle
  (n^2_{\chi,\bar\chi}  - n^{\rm eq \,2}_{\chi,\bar\chi})\,.
\end{eqnarray}
 In our work, we assumed $\langle \sigma_{\chi\chi} v\rangle=\langle \sigma_{\bar\chi\bar\chi} v\rangle$. Here $n_{\chi}$ and $n_{\bar\chi}$ are the number densities of ADM particles and anti-particles, with their equilibrium values expressed as:
\begin{eqnarray} \label{eq:B5}
n^{\rm eq}_{\chi} = g ~{\left( \frac{m T}{2 \pi} \right)}^{3/2}
  {\rm e}^{(-m + \mu)/T}\,,\,\,\,\,\,\,\,\,\,\,
  n^{\rm eq}_{\bar\chi} =  g ~{\left( \frac{m T}{2 \pi}
    \right)}^{3/2} {\rm e}^{(-m - \mu)/T}\,,
\end{eqnarray}
where $g$ is the number of internal degrees of freedom of the particle.

The Boltzmann Eq. (\ref{eq:B3}) can be rewritten in terms of the dimensionless quantity $Y = n/s$ and $x $, where the entropy density $s = (2\pi^2/45)g_{*s}T^3$ with $g_{*s}$ being the effective number of entropy degrees of freedom. The revised equation is
\begin{equation} \label{eq:B6}
  \frac{{\rm d} Y_{\chi,\bar\chi}}{{\rm d}x} =
  -\frac{\lambda }{x^2 A_{s,g}}
  \left[
      \langle \sigma_{\chi\bar\chi} v \rangle
      (Y_{\chi} Y_{\bar\chi} - Y^{\rm eq}_{\chi}Y^{\rm eq}_{\bar\chi}   )
      + \langle \sigma_{\chi\chi} v \rangle
    (Y^2_{\chi,\bar\chi} - Y^{\rm eq \,2}_{\chi,\bar\chi}   ) \right]\, ,
\end{equation}
where $A_{s}=\sqrt{ 1 + x_e^2/x^2 }$, $A_{g}=\left( x/x_t \right)^{2/3}$ and $\lambda=1.32m M_{\mathrm{Pl}}\sqrt{g_{*}}$.

At non-relativistic velocities, the expression for the thermal average of the annihilation cross-section multiplied by the relative velocity of ADM is
\begin{equation} \label{eq:B8}
   \langle \sigma v \rangle = a + 6\,b x^{-1} + {\cal O}(x^{-2})\, ,
\end{equation}
where $a$ and $b$ correspond to the $s$-wave and $p$-wave contributions to the annihilation cross section, respectively. During the early universe, ADM particles $\chi$ and their anti-particles $\bar{\chi}$ remain in thermal equilibrium. When the temperature decreases to the dark matter mass scale $T \lesssim m_\chi$ for $m_\chi > |\mu_\chi|$, the equilibrium number densities $n_{\chi,\mathrm{eq}}$ and $n_{\bar{\chi},\mathrm{eq}}$ experience exponential suppression. Subsequently, the Hubble expansion rate $H$ exceeds the annihilation rate, causing $\chi$ and $\bar\chi$ to decouple from thermal equilibrium. This results in the co-moving number densities remaining nearly constant over cosmic time evolution.

\subsection{The evolution of relic abundance of ADM}\label{sec:3}
To simplify Eq. (\ref{eq:B6}), we use the difference $\Delta_-$ and sum $\Delta_+$ of $Y_\chi$ and $Y_{\bar{\chi}}$
\begin{equation} \label{eq:C1}
  \Delta_- = Y_{\chi} - Y_{\bar\chi}\, ,\,\,\,\,\,
  \Delta_+ = Y_{\chi} + Y_{\bar\chi}\,,
\end{equation}
where $\Delta_-$ is the asymmetry factor. To ensure the survival of the ADM particle and anti-particle asymmetry, the self annihilation cross section must satisfy $\sigma_{\chi\chi} \ll \sigma_{\chi\bar{\chi}}$ to avoid complete wash-out caused by excessive self annihilation processes. In terms of $\Delta_-$ and $\Delta_+$, Eq. (\ref{eq:C1}) can be rewritten as
\begin{equation}\label{eq:C2}
  \frac{{\rm d} \Delta_-}{{\rm d}x} =
  -\frac{\lambda } {x^2 A_{s,g}}\,
  \langle \sigma_{\chi\chi} v \rangle
  (\Delta_+ \Delta_- - \Delta^{\rm eq}_+ \Delta^{\rm eq}_-);
\end{equation}
\begin{equation}\label{eq:C3}
  \frac{{\rm d} \Delta_+}{{\rm d}x} =
-\frac{1}{2}\,\frac{\lambda }{x^2 A_{s,g}}\,
  \langle \sigma_{\chi\bar\chi} v \rangle
  (\Delta^2_+- \Delta^2_- - {\Delta^{\rm eq}_+}^2 + {\Delta^{\rm eq}_-}^2)\,,
\end{equation}
where $\Delta^{\rm eq}_+ = 0.29\,g/g_* x^{3/2} {\rm e}^{-x}$ and $\Delta^{\rm eq}_- = 0.29\,g/g_* x^{3/2} {\rm e}^{-x}\,\mu/T$. From Eq. (\ref{eq:C2}), when the self annihilation cross section vanishes ($\sigma_{\chi\chi} = 0$), the asymmetry factor $\Delta_-$ remains constant \cite{Iminniyaz:2016iom}. The right-hand sides of Eqs. (\ref{eq:C2}) and (\ref{eq:C3}) include the self annihilation cross section $\sigma_{\chi\chi}$ and annihilation cross section $\sigma_{\chi\bar{\chi}}$ of ADM particles and anti-particles, respectively. In this work, we focus on studying the impact of self annihilation on the evolution of relic density of ADM particles and anti-particles. Therefore, we primarily analyze Eq. (\ref{eq:C2}) in the subsequent analysis.

During the phase from thermal equilibrium to the freeze-out temperature $x_{fw}$ for the self annihilation, $\Delta_+$ remains very close to the equilibrium value of $\Delta_+^{eq}$, and Eq. (\ref{eq:C2}) can be simplified to
\begin{equation}\label{eq:C4}
  \frac{{\rm d} \Delta_-}{{\rm d} x} =
    - \frac{\lambda }{x^2 A_{s,g}}\,
    \langle \sigma_{\chi\chi} v \rangle \Delta^{\rm eq}_+
    (\Delta_- - \Delta^{\rm eq}_-).
\end{equation}
By adopting the initial value of $\Delta_{\mathrm{-in}}$, which is close to the equilibrium value $\Delta_{-}^{eq}$ at temperatures $T\sim m$ (corresponding to $x\sim1$) and utilizing Eq. (\ref{eq:B8}), the above equation can be solved numerically, with additional parameters fixed at $\mu/T = 10^{-9}$ and $ x_{\text{in}}=1$.

\begin{figure}[h]
  \begin{center}
    \hspace*{-0.5cm} \includegraphics*[width=8cm]{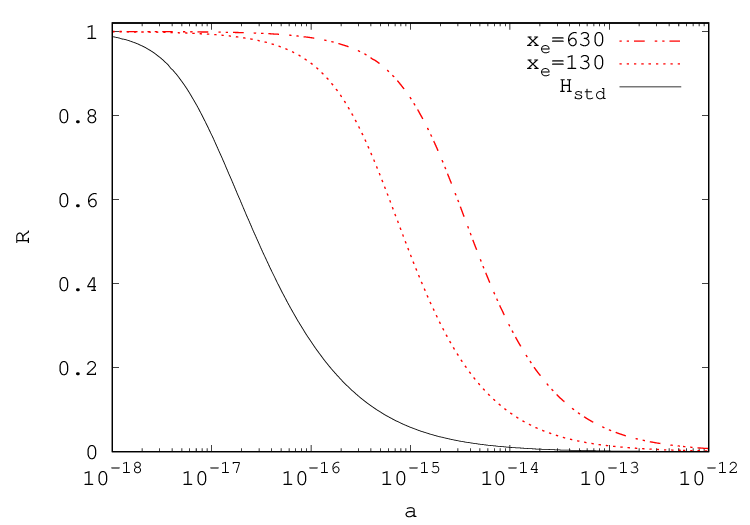}
    \put(-115,-12){(a)}
    \hspace*{-0.5cm} \includegraphics*[width=8cm]{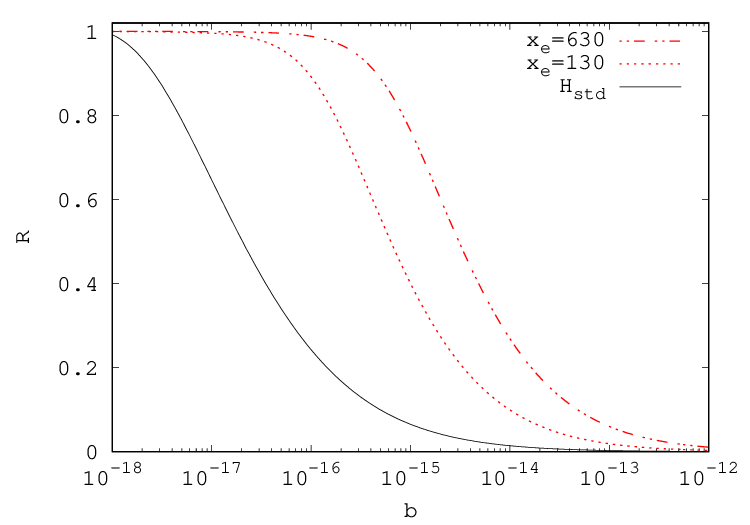}
    \put(-115,-12){(b)}
    \vspace{0.5cm}
     \hspace*{-0.5cm} \includegraphics*[width=8cm]{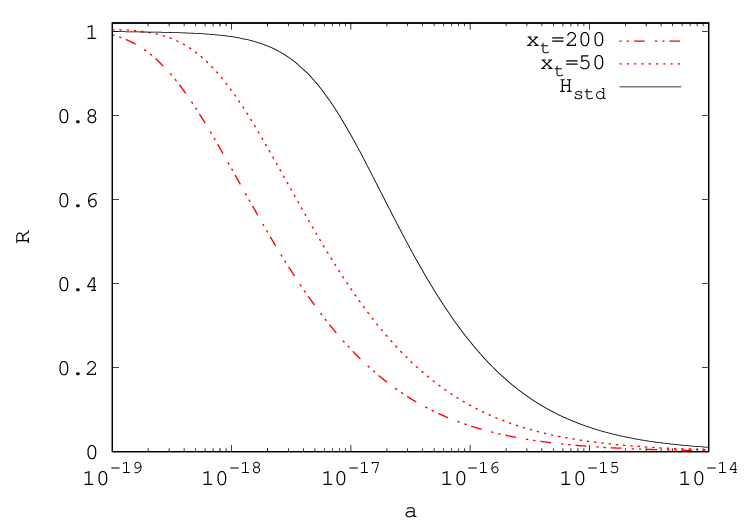}
    \put(-115,-12){(c)}
    \hspace*{-0.5cm} \includegraphics*[width=8cm]{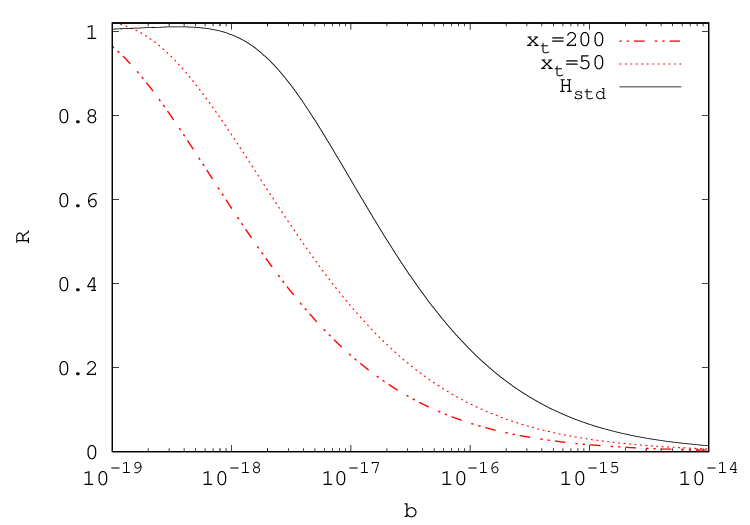}
    \put(-115,-12){(d)}
     \caption{\label{fig:R-a-b} \footnotesize
          The ratio of final asymmetry to the initial asymmetry $R = \Delta_{-\infty}/\Delta_{\mathrm{in}}$ as a function of $a$ and $b$ for the shear-dominated universe and Gauss-Bonnet braneworld. Here, (a) and (c) correspond to $s$-wave annihilation, while (b), (d) represent $p$-wave annihilation and $m=10\ \mathrm{GeV}$, $g=2$, $g_*=20$.   }
   \end{center}
\end{figure}
\begin{figure}[h]
  \begin{center}
    \hspace*{-0.5cm} \includegraphics*[width=8.7cm]{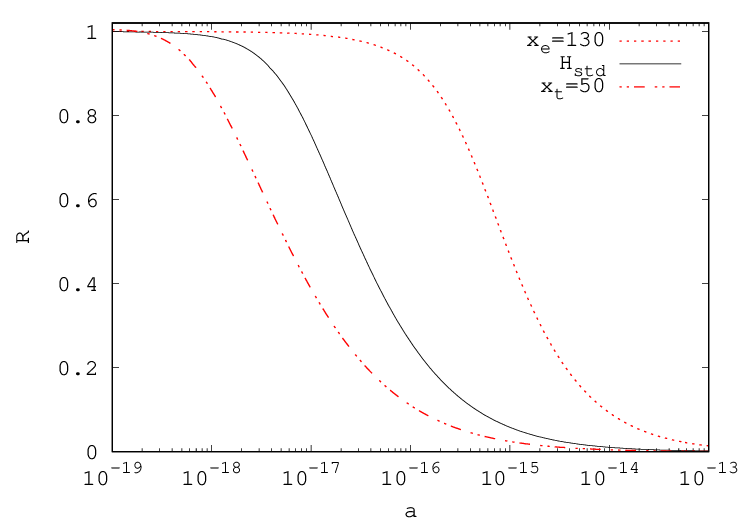}
    \caption{\label{fig:R-a}\footnotesize
    The ratio of final asymmetry to the initial asymmetry $R = \Delta_{-\infty}/\Delta_{\mathrm{in}}$ as a function of $a$ for the shear-dominated universe and Gauss-Bonnet braneworld. Here $m = 10$ GeV, $g = 2$, $g_* = 90$.}
     \end{center}
\end{figure}
\begin{figure}[h]
  \begin{center}
     \hspace*{-0.5cm} \includegraphics*[width=8cm]{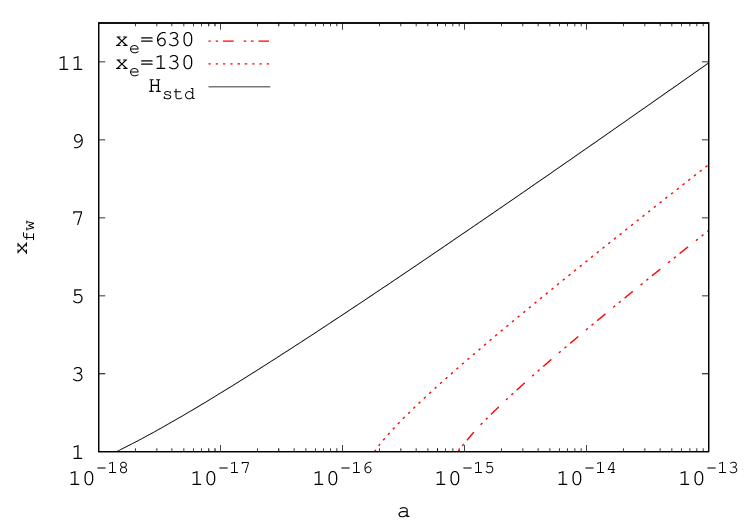}
    \put(-115,-12){(a)}
    \hspace*{-0.5cm} \includegraphics*[width=8cm]{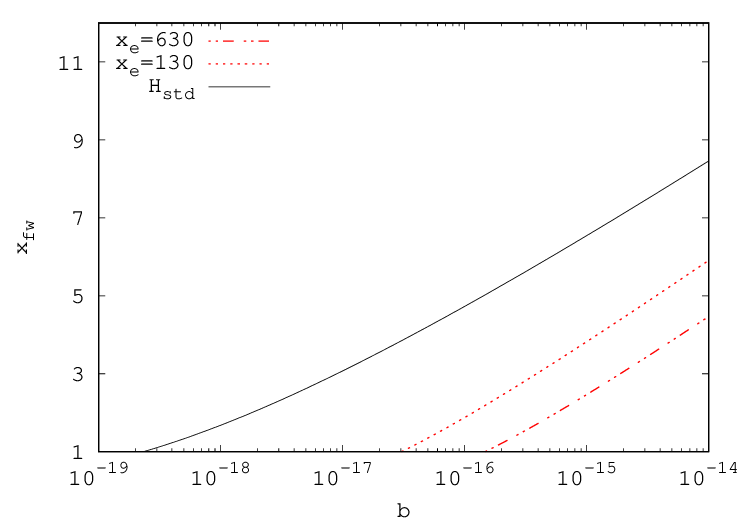}
    \put(-115,-12){(b)}
    \vspace{0.5cm}
     \hspace*{-0.5cm} \includegraphics*[width=8cm]{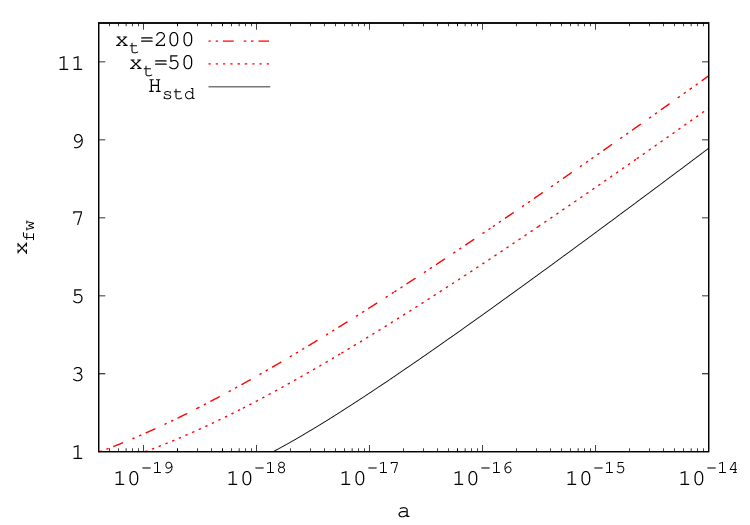}
    \put(-115,-12){(c)}
    \hspace*{-0.5cm} \includegraphics*[width=8cm]{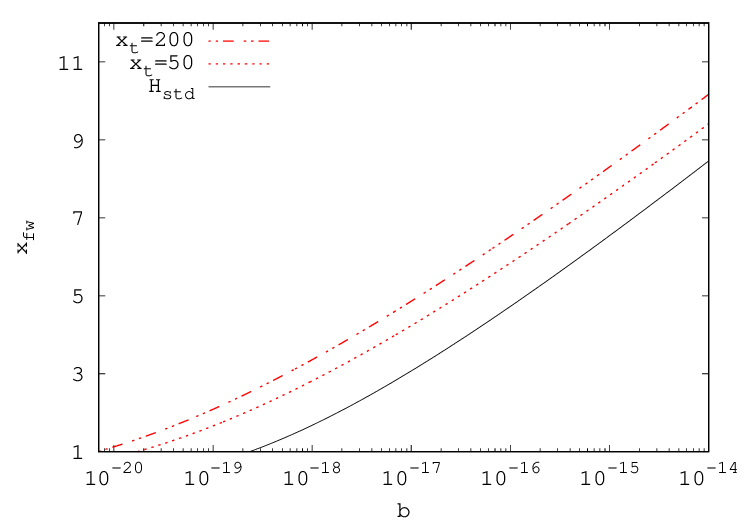}
    \put(-115,-12){(d)}
     \caption{\label{fig:xf-a} \footnotesize
             The inverse-scaled freeze-out point $x_{fw}$ as a function of $a$ and $b$. Here $m = 10$ GeV, $g = 2$, $g_* = 90$. }
   \end{center}
\end{figure}
From Eq. (\ref{eq:B8}) and above conditions, we present in Fig.~\ref{fig:R-a-b} the ratio of final asymmetry to the initial asymmetry $R \equiv \Delta_{-\infty}/\Delta_{\mathrm{-in}}$ as a function of the self annihilation cross section, where panels (a) and (b) correspond to results in the shear-dominated universe, and panels (c) and (d) correspond to those in the Gauss-Bonnet braneworld. In Fig.~\ref{fig:R-a-b}, (a) and (b) panels reveal that for both $s$-wave and $p$-wave annihilation cross section, the shear-dominated universe exhibits a higher upper limit of the self annihilation cross section compared to the standard cosmological scenario within the same $R$ range. Additionally, the upper limits on the self annihilation cross section increase with increasing the shear factor. For example, when $R > 0.1$, the upper limit in the standard case is $a \lesssim 4.37 \times 10^{-16}$; this value increases to $a \lesssim 8.84 \times 10^{-15}$ at $x_e=130$, and further rises to $a \lesssim 4.20 \times 10^{-14}$ when $x_e = 630$.

Similarly, from panels (c) and (d) in Fig.~\ref{fig:R-a-b}, we find that the Gauss-Bonnet braneworld exhibits lower upper limits of the self annihilation cross section compared to the standard cosmological scenario within the same $R$ range. Furthermore, the upper limits of the self annihilation cross section decrease with increasing $x_t$. For example, when $R > 0.1$, the upper limit under the standard scenario is $a \lesssim 4.37 \times 10^{-16}$; this value reduces to $a \lesssim 1.14 \times 10^{-16}$  at $x_t = 50$, and further declines to $a \lesssim 4.57 \times 10^{-17}$ for $x_t = 200$. In Fig.~\ref{fig:R-a}, we plot the relationship between $ R $ and the self annihilation cross section for shear-dominated universe with $ x_e=130 $, the standard cosmological scenario and Gauss-Bonnet braneworld with $ x_t=50 $. From Fig.~\ref{fig:R-a}, it can be found that the shear-dominated universe and Gauss-Bonnet braneworld correspond to the enhanced and weakened Hubble expansion rates, respectively. Therefore, the conclusions on the upper limit of the self annihilation cross section for the same $R$ range are opposite for both scenarios.

By adopting the methods for solving the approximate solutions of the freeze-out temperature $x_{fw}$ of wash-out asymmetry process from Ref.\cite{Liu:2023pfd}, the approximate solution of $x_{fw}$ in the shear-dominated universe and Gauss-Bonnet braneworld can be derived as
\begin{equation}\label{eq:C9}
 x_{fw}=\ln\frac{0.29\,\xi\lambda g\langle\sigma_{\chi\chi}v\rangle}{g_*x^{1/2}A_{s,g}} \  \Bigg|_{x=x_{fw}}\,.
\end{equation}
where $\xi$ is the numerical constant of order unity, we use $\xi=1$.
From the above equation, we present in Fig.~\ref{fig:xf-a} the relationship between the freeze-out temperature $x_{fw}$ and self annihilation cross section. Based on the analysis, it is found that both for $s$-waves and $p$-waves, within a given range of $x_{fw}$ values, the shear-dominated universe exhibits a higher upper limit on the self annihilation cross section compared to the standard model, while the Gauss-Bonnet braneworld scenario shows a relatively smaller upper limit. We obtain the approximate solution of Eq. (\ref{eq:C4}) as
\begin{equation}\label{eq:C10}
 R_{\rm app}(x_{fw}) \equiv \frac{\Delta^{\rm eq}_-(x_{fw})}{\Delta^{\rm eq}_-(1)}\approx \frac{x_{fw}^{3/2}e^{-x_{fw}}}{e^{-1}}\,.
\end{equation}
Having determined the numerical and approximate solutions for Eq. (\ref{eq:C4}), we now compare the values of $R$ and $ R_{\text{app}}$. Under the condition $ R = 0.1 $, the freeze-out points $ x_{\text{fw}} $  of wash-out asymmetry process for the three cosmological scenarios are found as follows: $ x_{\text{fw}} = 5.86 $ at $ a = 4.37 \times 10^{-16} $ in the standard cosmology, $ x_{\text{fw}} = 5.75 $ at $ a = 8.84 \times 10^{-15} $ in the shear-dominated universe ($ x_e = 130 $), and $ x_{\text{fw}} = 5.93 $ at $ a = 1.14 \times 10^{-16} $ in the Gauss-Bonnet braneworld ($ x_t = 50 $). Substituting these into Eq. (\ref{eq:C10}) yields $ R_{\text{app}} = 0.11 $, $ 0.12 $, and $ 0.10 $ respectively, demonstrating strong consistency between numerical and approximate solutions of Eq. (\ref{eq:C4}).

The analytical approximate solutions Eq. (\ref{eq:C9}) and Eq. (\ref{eq:C10}) can be used to analyze the constraints imposed by asymmetry on the self-annihilation cross section.
These two equations respectively describe the functional relations between the freeze-out temperature and cross section, the asymmetry ratio and freeze-out temperature. When combined, they establish the relationship between asymmetry ratio $R$ and cross section $\left\langle\sigma v\right\rangle$. For example, when we choose $R=0.1$ as the lower limit, the freeze-out point derived from Eq. (\ref{eq:C10}) is $x_{fw}=5.99$. Plugging this value $x_{fw}=5.99$ into Eq. (\ref{eq:C9}), we can obtain the corresponding self-annihilation cross section. When $R=0.1$, we fnd that the Eqs. (\ref{eq:C9}) and (\ref{eq:C10}) require that constraint on the $s$-wave cross section $a$ and the constraint on $p$-wave $6b/x$ must be same. The above discussion is independent of concrete ADM models and cosmological models, provided that the dark matter thermal equilibrium assumption is valid. Therefore, this perspective could also be utilized to compare and test some ADM particle models.

\subsection{The constraints on the winos mass}\label{sec:4}
\begin{figure}[h]
  \begin{center}
     \hspace*{-0.5cm} \includegraphics*[width=8cm]{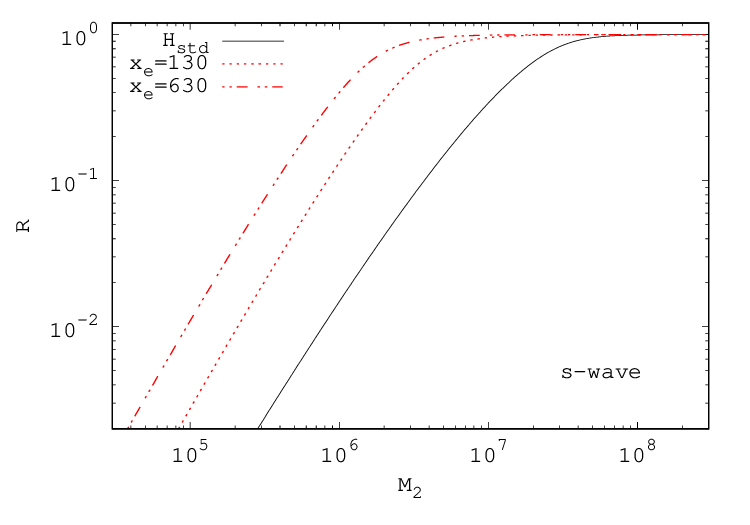}
    \put(-115,-12){(a)}
    \hspace*{-0.5cm} \includegraphics*[width=8cm]{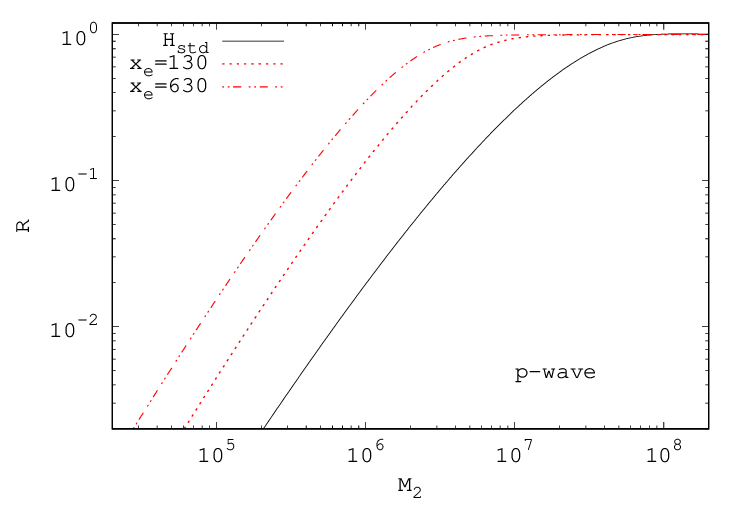}
    \put(-115,-12){(b)}
    \vspace{0.5cm}
     \hspace*{-0.5cm} \includegraphics*[width=8cm]{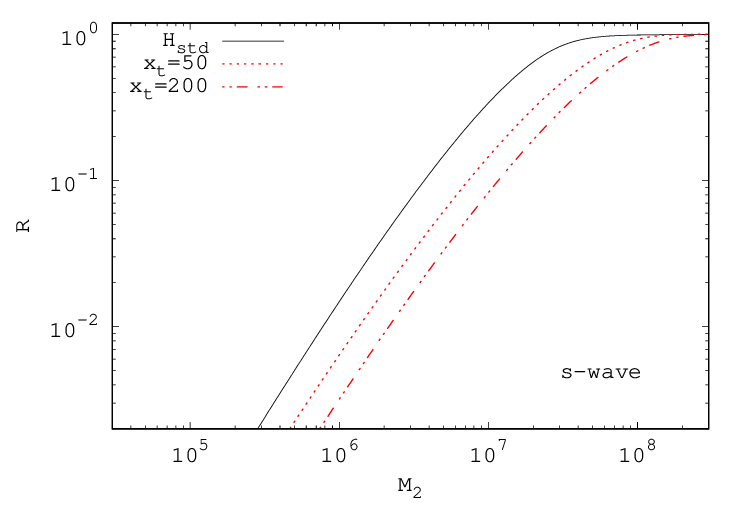}
    \put(-115,-12){(c)}
    \hspace*{-0.5cm} \includegraphics*[width=8cm]{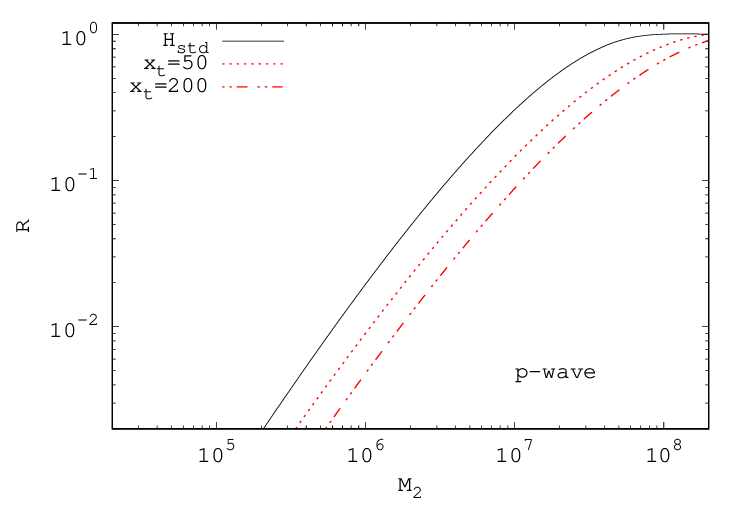}
    \put(-115,-12){(d)}
     \caption{\label{fig:R-M} \footnotesize
             The ratio of final asymmetry to the initial asymmetry $R = \Delta_{-\infty}/\Delta_{\mathrm{in}}$ as a function of the wino mass $M_2$. Here $m = 10$ GeV, $g = 2$, $g_* = 90$.}
   \end{center}
\end{figure}
When considering the left-handed sneutrino as ADM candidate in supersymmetric models \cite{Hooper:2004dc,Abel:2006nv,Kang:2011cni}, the calculation of its self annihilation cross section $\sigma_{\chi\chi}$ requires the inclusion of the $t$-channel exchange mechanism mediated by electroweak gauginos. Electroweak gauginos include binos with mass $M_1$ and winos with mass $M_2$. The gaugino masses are much larger than the sneutrino mass, the expression for the self annihilation cross section is
\begin{equation}\label{eq:D1}
\langle\sigma_{\tilde{\nu}_L \tilde{\nu}_L} v\rangle \simeq \frac{g_2^4}{16\pi}(1-\frac{3}{2x})(\frac{\tan^2 {\theta_W}}{M_1}+\frac{1}{M_2})^2\ ,
\end{equation}
where $g_2=e/\sin\theta_W$ is $SU(2)$ gauge coupling and $\theta_W$ is the weak mixing angle, $e$ is the charge of electron. Similar to sneutrinos, higgsinos exhibit identical couplings to electroweak gauginos. Through $t$-channel exchanges involving bino and wino, these higgsinos undergo self-annihilation into $H_u$ and $H_d$. This scattering process corresponds to $p$-wave interaction, and following the analysis in \cite{Nihei:2002ij}, the self annihilation cross section is
\begin{equation}\label{eq:D2}
\left\langle\sigma_{\tilde{h}_u\tilde{h}_u} v\right\rangle \simeq \frac{3g_2^4}{8\pi x}\left( \frac{\tan^2\theta_w}{M_1} + \frac{1}{M_2} \right)^2.
\end{equation}
In Eq. (\ref{eq:D1}) and Eq. (\ref{eq:D2}), we have presented the expressions for the self-annihilation cross sections in the $s$-wave and $p$-wave scenarios, respectively. Further, by taking into account the conditions $M_1\simeq M_2/2$, $\tan^2\theta_W\approx0.3$, and Eq. (\ref{eq:C4}), we present the function relationship between the wino mass $M_2$ and $R$ in Fig.~\ref{fig:R-M}. From Fig.~\ref{fig:R-M}, we can find that, both for the $s$-wave and $p$-wave cases, when the standard model, shear-dominated universe, and Gauss-Bonnet braneworld are under the same asymmetric conditions, $M_2$ has a smaller lower limit in the shear-dominated universe compared to the standard model, while the Gauss-Bonnet braneworld has a higher lower limit. For example, when $R > 0.1$, we consider the $s$-wave in the standard model, shear-dominated universe and Gauss-Bonnet braneworld scenarios, then obtain
\begin{equation}
\begin{aligned}
&M_2 \gsim 8.3\times10^5 \,{\rm GeV},\ x_e = 130, \\
&M_2 \gsim 3.8\times10^5 \,{\rm GeV},\ x_e = 630, \\
&M_2 \gsim 7.2\times10^6 \,{\rm GeV},\ x_t = 50, \\
&M_2 \gsim 1.1\times10^7 \,{\rm GeV},\ x_t = 200, \\
&M_2 \gsim 3.5\times10^6 \,{\rm GeV}\,({\rm standard\ case}).
\end{aligned}
\end{equation}
For the $p$-wave case under $R > 0.1$, we can obtain
\begin{equation}
\begin{aligned}
&M_2 \gsim 8.0\times10^5 \,{\rm GeV},\ x_e = 130, \\
&M_2 \gsim 3.5\times10^5 \,{\rm GeV},\ x_e = 630, \\
&M_2 \gsim 6.9\times10^6 \,{\rm GeV},\ x_t = 50, \\
&M_2 \gsim 1.1\times10^7 \,{\rm GeV},\ x_t = 200, \\
&M_2 \gsim 3.5\times10^6 \,{\rm GeV}\,({\rm standard\ case}).
\end{aligned}
\end{equation}

\section{Summary and conclusions}\label{sec:5}
In this paper, we first reviewed the Hubble expansion rate in the shear-dominated universe and Gauss-Bonnet braneworld, then substituted it into the Boltzmann equation for ADM particles and anti-particles which include the self annihilation process. Through further calculations and simplifications, we ultimately obtained the evolution Eq. (\ref{eq:C4}) of the relic density of ADM.

We solved the evolution equation of the asymmetry using both numerical and approximate analytical methods. In the numerical approach, the evolution equation of the asymmetry was solved to obtain the $ R = \Delta_{-\infty}/\Delta_{\text{in}} $ as a function of the self annihilation cross section. The results revealed that, compared to the standard cosmological scenario within the same $ R $ range, the shear-dominated universe exhibits a higher upper limit on the self annihilation cross section, while the Gauss-Bonnet braneworld shows a lower upper limit.

For the approximate solution, we adopt the method of solving the approximate solution of the freeze-out temperature $ x_{\text{fw}} $ of wash-out asymmetry process in \cite{Liu:2023pfd}, and present the expression of $ x_{\text{fw}} $ in the shear-dominated universe and Gauss-Bonnet braneworld. Finally, by comparing the values of $ R $ and $ R_{\text{app}} $, we verified that the numerical solutions and approximate evolution equation of the relic density are in close agreement.

We investigated the constraints on the wino mass $ M_2 $ for sneutrino and higgsino ADM in the shear-dominated universe and the Gauss-Bonnet braneworld. Under the condition $ M_1, M_2 \gg m $, using Eq. (\ref{eq:D2}) for the self annihilation cross section and the evolution equation of the asymmetry, we plotted the relationship between $ M_2 $ and $ R $ for these scenarios in Fig.~\ref{fig:R-M}. Our analysis shows that when the standard model, shear-dominated universe, and Gauss-Bonnet braneworld are compared within the same range of $ R $, the lower limit on $ M_2 $ is smaller in the shear-dominated universe and higher in the Gauss-Bonnet braneworld relative to the standard model.

The above results reveal that in the shear-dominated universe and Gauss-Bonnet braneworld, constraints must be imposed on the self-annihilation cross section during ADM self annihilation processes to ensure that the final ADM maintains a considerable final asymmetry. This constraint further establishes new limiting boundaries for the wino mass $M_2$ range of sneutrino and higgsino ADM candidates in non-standard cosmologies.

\section*{Acknowledgments}

The work is supported by the National Natural Science Foundation of China (Grant No. 12463001).


\begin{thebibliography}{99}
\bibitem{Planck:2018vyg}
N.~Aghanim \textit{et al.} [Planck],
Astron. Astrophys. \textbf{641}, A6 (2020)
[erratum: Astron. Astrophys. \textbf{652}, C4 (2021)],
doi:10.1051/0004-6361/201833910,
arXiv:1807.06209 [astro-ph.CO].

\bibitem{Pato:2015dua}
M.~Pato, F.~Iocco and G.~Bertone,
J. Cosmol. Astropart. Phys. \textbf{12}, 001 (2015),
doi:10.1088/1475-7516/2015/12/001,
arXiv:1504.06324 [astro-ph.GA].

\bibitem{Iocco:2015xga}
F.~Iocco, M.~Pato and G.~Bertone,
Nature Phys. \textbf{11}, 245-248 (2015),
doi:10.1038/nphys3237,
arXiv:1502.03821 [astro-ph.GA].

\bibitem{Jiao:2023aci}
Y.~Jiao, F.~Hammer, H.~Wang, J.~Wang, P.~Amram, L.~Chemin and Y.~Yang,
Astron. Astrophys. \textbf{678}, A208 (2023),
doi:10.1051/0004-6361/202347513,
arXiv:2309.00048 [astro-ph.GA].

\bibitem{Ou:2023adg}
X.~Ou, A.~C.~Eilers, L.~Necib and A.~Frebel,
Mon. Not. Roy. Astron. Soc. \textbf{528}, no.1, 693-710 (2024),
doi:10.1093/mnras/stae034,
arXiv:2303.12838 [astro-ph.GA].

\bibitem{Clowe:2006eq}
D.~Clowe, M.~Bradac, A.~H.~Gonzalez, M.~Markevitch, S.~W.~Randall, C.~Jones and D.~Zaritsky,
Astrophys. J. Lett. \textbf{648}, L109-L113 (2006),
doi:10.1086/508162,
arXiv:astro-ph/0608407 [astro-ph].

\bibitem{Robertson:2016qef}
A.~Robertson, R.~Massey and V.~Eke,
Mon. Not. Roy. Astron. Soc. \textbf{467}, no.4, 4719-4730 (2017),
doi:10.1093/mnras/stx463,
arXiv:1612.03906 [astro-ph.CO].

\bibitem{Springel:2005nw}
V.~Springel, S.~D.~M.~White, A.~Jenkins, C.~S.~Frenk, N.~Yoshida, L.~Gao, J.~Navarro, R.~Thacker, D.~Croton and J.~Helly, \textit{et al.}
Nature \textbf{435}, 629-636 (2005),
doi:10.1038/nature03597,
arXiv:astro-ph/0504097 [astro-ph].

\bibitem{Bertone:2004pz}
G.~Bertone, D.~Hooper and J.~Silk,
Phys. Rept. \textbf{405}, 279-390 (2005),
doi:10.1016/j.physrep.2004.08.031,
arXiv:hep-ph/0404175 [hep-ph].

\bibitem{Nussinov:1985xr}
S.~Nussinov,
Phys. Lett. B \textbf{165}, 55-58 (1985),
doi:10.1016/0370-2693(85)90689-6.

\bibitem{Chivukula:1989qb}
R.~S.~Chivukula and T.~P.~Walker,
Nucl. Phys. B \textbf{329}, 445-463 (1990),
doi:10.1016/0550-3213(90)90151-3.

\bibitem{An:2009vq}
H.~An, S.~L.~Chen, R.~N.~Mohapatra and Y.~Zhang,
J. High Energy Phys. \textbf{03}, 124 (2010),
doi:10.1007/JHEP03(2010)124,
arXiv:0911.4463 [hep-ph].

\bibitem{Cohen:2009fz}
T.~Cohen and K.~M.~Zurek,
Phys. Rev. Lett. \textbf{104}, 101301 (2010),
doi:10.1103/PhysRevLett.104.101301,
arXiv:0909.2035 [hep-ph].

\bibitem{Shelton:2010ta}
J.~Shelton and K.~M.~Zurek,
Phys. Rev. D \textbf{82}, 123512 (2010),
doi:10.1103/PhysRevD.82.123512,
arXiv:1008.1997 [hep-ph].

\bibitem{Hooper:2004dc}
D.~Hooper, J.~March-Russell and S.~M.~West,
Phys. Lett. B \textbf{605}, 228-236 (2005),
doi:10.1016/j.physletb.2004.11.047,
arXiv:hep-ph/0410114 [hep-ph].

\bibitem{Abel:2006nv}
S.~Abel and V.~Page,
AIP Conf. Proc. \textbf{878}, no.1, 341-346 (2006),
doi:10.1063/1.2409106,
arXiv:hep-ph/0609140 [hep-ph].

\bibitem{Kaplan:2009ag}
D.~E.~Kaplan, M.~A.~Luty and K.~M.~Zurek,
Phys. Rev. D \textbf{79}, 115016 (2009),
doi:10.1103/PhysRevD.79.115016,
arXiv:0901.4117 [hep-ph].

\bibitem{XENON100:2011uwh}
E.~Aprile \textit{et al.} [XENON100],
Phys. Rev. Lett. \textbf{107}, 131302 (2011),
doi:10.1103/PhysRevLett.107.131302,
arXiv:1104.2549 [astro-ph.CO].

\bibitem{Kolb}
Kolb, Edward W., and Michael S. Turner,
Addison-Wesley (1990).


\bibitem{Ellwanger:2012yg}
U.~Ellwanger and P.~Mitropoulos,
J. Cosmol. Astropart. Phys. \textbf{07}, 024 (2012),
doi:10.1088/1475-7516/2012/07/024,
arXiv:1205.0673 [hep-ph].

\bibitem{Liu:2023pfd}
F.~Liu and H.~Iminniyaz,
Chin. J. Phys. \textbf{88}, 380-387 (2024),
doi:10.1016/j.cjph.2024.02.007,
arXiv:2309.09155 [hep-ph].

\bibitem{Iminniyaz:2013cla}
H.~Iminniyaz and X.~Chen,
Astropart. Phys. \textbf{54}, 125-131 (2014),
doi:10.1016/j.astropartphys.2013.12.003,
arXiv:1308.0353 [hep-ph].

\bibitem{Meehan:2014zsa}
M.~T.~Meehan and I.~B.~Whittingham,
J. Cosmol. Astropart. Phys. \textbf{06}, 018 (2014),
doi:10.1088/1475-7516/2014/06/018,
arXiv:1403.6934 [astro-ph.CO].

\bibitem{Abdusattar:2015azp}
H.~Abdusattar and H.~Iminniyaz,
Commun. Theor. Phys. \textbf{66}, no.3, 363-368 (2016),
doi:10.1088/0253-6102/66/3/363,
arXiv:1505.03716 [hep-ph].

\bibitem{Meehan:2015cna}
M.~T.~Meehan and I.~B.~Whittingham,
J. Cosmol. Astropart. Phys. \textbf{12}, 011 (2015),
doi:10.1088/1475-7516/2015/12/011,
arXiv:1508.05174 [astro-ph.CO].

\bibitem{Barrow:1982ei}
J.~D.~Barrow,
Nucl. Phys. B \textbf{208}, 501-508 (1982),
doi:10.1016/0550-3213(82)90233-4.

\bibitem{Kamionkowski:1990ni}
M.~Kamionkowski and M.~S.~Turner,
Phys. Rev. D \textbf{42}, 3310-3320 (1990),
doi:10.1103/PhysRevD.42.3310.

\bibitem{Iminniyaz:2016iom}
H.~Iminniyaz,
Phys. Lett. B \textbf{765}, 6-10 (2017),
doi:10.1016/j.physletb.2016.11.006,
arXiv:1604.04251 [hep-ph].

\bibitem{Langlois:2002bb}
D.~Langlois,
Prog. Theor. Phys. Suppl. \textbf{148}, 181-212 (2003),
doi:10.1143/PTPS.148.181,
arXiv:hep-th/0209261 [hep-th].

\bibitem{Okada:2009xe}
N.~Okada and S.~Okada,
Phys. Rev. D \textbf{79}, 103528 (2009),
doi:10.1103/PhysRevD.79.103528,
arXiv:0903.2384 [hep-ph].

\bibitem{Meehan:2014bya}
M.~T.~Meehan and I.~B.~Whittingham,
J. Cosmol. Astropart. Phys. \textbf{12}, 034 (2014),
doi:10.1088/1475-7516/2014/12/034,
arXiv:1404.4424 [astro-ph.CO].

\bibitem{Kang:2011cni}
Z.~Kang, J.~Li, T.~Li, T.~Liu and J.~M.~Yang,
Eur. Phys. J. C \textbf{76}, no.5, 270 (2016),
doi:10.1140/epjc/s10052-016-4114-9,
arXiv:1102.5644 [hep-ph].

\bibitem{Nihei:2002ij}
T.~Nihei, L.~Roszkowski and R.~Ruiz de Austri,
J. High Energy Phys. \textbf{03}, 031 (2002),
doi:10.1088/1126-6708/2002/03/031,
arXiv:hep-ph/0202009 [hep-ph].



\end{thebibliography}
\end{document}